\documentclass[aps,secnumarabic,nobalancelastpage,
nofootinbib,onecolumn,12pt]{revtex4}%
\usepackage{amsfonts}
\usepackage{amsmath}
\usepackage{graphics}
\usepackage{graphicx}
\usepackage{longtable}
\usepackage{url}
\usepackage{bm}%
\usepackage{amssymb}
\begin{document}
\title{Poisson-Fermi Modeling of Ion Activities in Aqueous Single and Mixed
Electrolyte Solutions at Variable Temperature}
\author{Jinn-Liang Liu}
\affiliation{Institute of Computational and Modeling Science, National Tsing Hua
University, Hsinchu 300, Taiwan. E-mail: jlliu@mx.nthu.edu.tw}
\author{Bob Eisenberg}
\affiliation{Department of Applied Mathematics, Illinois Institute of Technology, Chicago
IL 60616. USA}
\date{\today }

\begin{abstract}

\end{abstract}
\maketitle

\textbf{Abstract.} The combinatorial explosion of empirical parameters in tens
of thousands presents a tremendous challenge for extended Debye-H\"{u}ckel
models to calculate activity coefficients of aqueous mixtures of most
important salts in chemistry. The explosion of parameters originates from the
phenomenological extension of the Debye-H\"{u}ckel theory that does not take
steric and correlation effects of ions and water into account. In contrast,
the Poisson-Fermi theory developed in recent years treats ions and water
molecules as nonuniform hard spheres of any size with interstitial voids and
includes ion-water and ion-ion correlations. We present a Poisson-Fermi model
and numerical methods for calculating the individual or mean activity
coefficient of electrolyte solutions with any arbitrary number of ionic
species in a large range of salt concentrations and temperatures. For each
activity-concentration curve, we show that the Poisson-Fermi model requires
only three unchanging parameters at most to well fit the corresponding
experimental data. The three parameters are associated with the Born radius of
the solvation energy of an ion in electrolyte solution that changes with salt
concentrations in a highly nonlinear manner.

\section{Introduction}

Thermodynamic modeling of aqueous electrolyte solutions plays an important
role in chemical and biological sciences
\cite{RS59,N91,P95,H01,LM03,F04,LJ08,KF09,K10,VW16,V11,E13,RK15}. Despite
intense efforts in the past century, robust thermodynamic modeling of
electrolyte solutions still presents a difficult challenge and remains a
remote ambition in the extended Debye-H\"{u}ckel (DH) models due to the
enormous number of parameters that need to be adjusted, carefully and often
subjectively \cite{V11,RK15}. For example, the Pitzer model requires 8
parameters for a ternary system and up to 8 temperature coefficients
(parameters) for every Pitzer parameter in a temperature interval from 0 to
about 200 $%
\operatorname{{}^{\circ}{\rm C}}%
$ \cite{V11,RK15}. It is indeed a frustrating despair (\textit{frustration} on
p. 11 in \cite{K10} and \textit{despair }on p. 301 in \cite{RS59}) that
approximately 22,000 parameters for combinatorial solutions of the most
important 28 cations and 16 anions in salt chemistry have to be extracted from
the available experimental data for one temperature \cite{V11}. The Pitzer
model is still the most widely used DH model with unmatched precision for
modeling aqueous electrolyte solutions over wide ranges of composition,
temperature, and pressure \cite{RK15}.

The Pitzer model and its variants \cite{RK15} are all derived from the
Debye-H\"{u}ckel theory \cite{DH23} that in turn is based on a linear
Poisson-Boltzmann (PB) equation \cite{LM03} although potentials calculated
from PB near ions (for example) are often far beyond the linear range of the
potential near ions or interfaces. The PB equation treats ions as point
charges without steric volumes and water molecules as a homogeneous dielectric
medium without steric volumes either and with a constant dielectric constant
that neglects ion-water and ion-ion correlations. These simplifications give
rise to the elegant, simple, and useful DH theory. However, it is precisely
because of the linearization and simplifications on steric and correlation
effects that extended DH models have needed an explosion in the number of
parameters in order to overcome the deficiencies (simplifications) of the
classical Poisson-Boltzmann theory. The nonlinear PB equation was developed by
Gouy and Chapman \cite{G10,C13}.

In the past few years, we have intensively investigated these two effects in a
range of areas from electric double layers \cite{L13,LX17}, ion activities
\cite{LE15a}, to biological ion channels \cite{LE13,LX17,LE14,LE14a,LE15,LH16}
and consequently developed an advanced theory --- the Poisson-Fermi (PF)
theory --- that treats ions and water molecules as nonuniform hard spheres of
any size with interstitial voids and includes many of the correlation effects
of ions and water. We refer to our previous papers and references therein for
a historical account of the literature of this theory. In \cite{LE15a}, we
proposed a PF model for calculating activity coefficients of individual ions
in aqueous single NaCl and CaCl$_{2}$ electrolyte solutions at the temperature
298.15 K. The model is further tested in this paper for eight 1:1 electrolytes
(LiCl, LiBr, NaF, NaCl, NaBr, KF, KCl, and KBr), six 2:1 electrolytes
(MgCl$_{2}$, MgBr$_{2}$, CaCl$_{2}$, CaBr$_{2}$, BaCl$_{2}$, and BaBr$_{2}$),
one mixed electrolyte (NaCl + MgCl$_{2}$), one 1:1 electrolyte (NaCl ) at
various temperatures from 298.15 to 573.15 K, and one 2:1 electrolyte
(MgCl$_{2}$) at various temperatures from 298.15 to 523.15 K, for which the
experimental data were compiled by Valisk\'{o} and Boda in \cite{VB15} and
Rowland et al. in \cite{RK15} from various experimental sources in
\cite{WR04,PP84,BH84,A92,GN78,WP98,C09,L65,KZ24,T21}.

The PF model is developed to calculate individual ion activities for which
experimental measurements and determination \cite{VW16,WV05,WR06},
interpretation of measurement data \cite{WR04,WR06,WV03,WV11}, and comparison
of different experimental methods \cite{WR06,WV16} have been extensively
investigated by Wilczek-Vera, Rodil, and Vera in the past two decades. PF
results on mean activity coefficients can be compared with experimental
measurements using the Debye-H\"{u}ckel equation of individual ion activities
\cite{LM03}.

In contrast to the Pitzer model, we show that all experimental data sets of
individual or mean activity coefficients as a function of variable
concentration in single electrolytes or mixtures at various temperatures can
be well fitted by the PF model with only 3 parameters at most for each
activity-concentration data curve. The model is characterized by three
different domains, namely, the Born ion, hydration shell, and remaining
solvent domains in which the Born ion domain is most crucial because all
activities around an ion are mainly governed by the singular charge of the ion
located at the center of the domain. The Born ion domain is defined by the
Born radius of the solvated ion, which is unknown and changes with salt
concentrations in a highly nonlinear manner.

The three parameters characterize three orders of approximation of the Born
radius in terms of ionic concentrations. Parameter 1 describes a correction of
the experimental Born radius of a single ion in pure water without any other
ions. Parameter 2 describes an adjustment of the unknown Born radius in
electrolyte solution that accounts for the Debye screening effect, which is
proportional to the square root of the ionic strength of the solution.
Parameter 3 is an adjustment in the next order approximation beyond the DH
treatment of ionic atmosphere. The physical origin of these parameters is
clear unlike that of most parameters in the Pitzer method \cite{F10,V11}. It
may even be possible in later work to calculate some of these parameters from
more detailed versions of our model.

Our approach to partition the free energy domain of a solvated ion into the
above three sub-domains yields a better approximation to calculate the free
energy since these sub-domains are determined by the experimental data of
solvation and thus separate short- and long-range interactions of the ion in a
more accurate way. This approach nevertheless incurs more complicated
numerical methods for solving the nonlinear partial differential equations of
the PF model in different domains with suitable interface conditions
\cite{L13}. We therefore present numerical methods in detail for future
verification and development of the present work.

\section{Theory}

\begin{figure}[t]
\centering\includegraphics[scale=0.7]{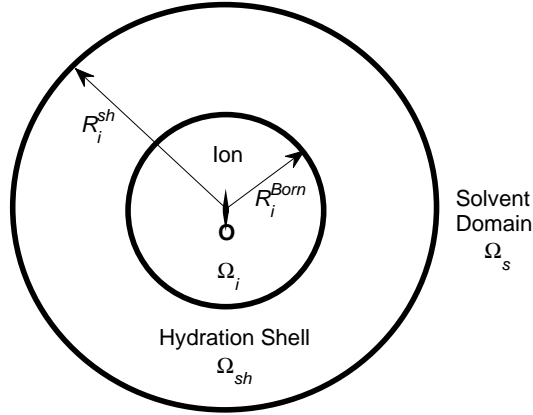}\caption{ The model domain
$\Omega$ is partitioned into the ion domain $\Omega_{i}$ (with radius
$R_{i}^{Born}$), the hydration shell domain $\Omega_{sh}$ (with radius
$R_{i}^{sh}$), and the remaining solvent domain $\Omega_{s}$.}%
\end{figure}

For an aqueous electrolyte solution with $K$ species of ions, the
Poisson-Fermi theory proposed in \cite{LX17,LE14} treats all ions and water of
any diameter as nonuniform hard spheres with interstitial voids between these
spheres. The activity coefficient $\gamma_{i}$ of an ion of species $i $ in
the solution describes the deviation of the chemical potential of the ion from
ideality ($\gamma_{i}=1$). The excess chemical potential $\mu_{i}^{\text{ex}%
}=k_{B}T\ln\gamma_{i}$ can be calculated by \cite{BC00,LE15a}%
\begin{equation}
\mu_{i}^{\text{ex}}=\Delta G_{i}-\Delta G_{i}^{0}\text{, \ }\Delta G_{i}%
=\frac{1}{2}q_{i}\phi(\mathbf{0})\text{, \ }\Delta G_{i}^{0}=\frac{1}{2}%
q_{i}\phi^{0}(\mathbf{0})\text{,}\tag{1}%
\end{equation}
where $k_{B}$ is the Boltzmann constant, $T$ is an absolute temperature,
$q_{i}$ is the ionic charge of the hydrated ion (also denoted by $i$),
$\phi(\mathbf{r})$ is a potential function of spatial variable $\mathbf{r}$ in
the domain $\Omega$ $=$ $\overline{\Omega}_{i}\cup\overline{\Omega}_{sh}%
\cup\Omega_{s}$ shown in Fig. 1, $\Omega_{i}$\ is the spherical domain
occupied by the ion $i$, $\Omega_{sh}$ is the hydration shell domain of the
ion, $\Omega_{s}$ is the remaining solvent domain, $\mathbf{0}$ denotes the
center (set to the origin) of the ion, $\phi(\mathbf{0})$ is the value of
$\phi(\mathbf{r})$ at $\mathbf{r}=\mathbf{0}$, and $\phi^{0}(\mathbf{r})$ is a
potential function when the solvent domain $\Omega_{s}$ does not contain any
ions at all with pure water only. The potential function $\phi(\mathbf{r})$
can be found by solving the Poisson-Fermi equation \cite{LX17}%
\begin{equation}
\left(  l_{c}^{2}\nabla^{2}-1\right)  \nabla\cdot\epsilon(\mathbf{r}%
)\nabla\phi(\mathbf{r})=\rho(\mathbf{r})\text{,}\tag{2}%
\end{equation}%
\begin{align}
\epsilon(\mathbf{r}) &  =\left\{
\begin{array}
[c]{l}%
\epsilon_{s}=\epsilon_{\text{w}}\epsilon_{0}\text{ in }\overline{\Omega}%
_{sh}\cup\Omega_{s}\\
\epsilon_{i}=\epsilon_{\text{ion}}\epsilon_{0}\text{ in }\overline{\Omega}_{i}%
\end{array}
\right.  ,\text{ }l_{c}=\left\{
\begin{array}
[c]{l}%
2a_{j}\text{ in }\overline{\Omega}_{sh}\cup\Omega_{s}\\
0\text{ in }\overline{\Omega}_{i}%
\end{array}
\right.  ,\tag{3}\\
\rho(\mathbf{r}) &  =\left\{
\begin{array}
[c]{l}%
\rho_{s}(\mathbf{r})=\sum_{k=1}^{K}q_{k}C_{k}(\mathbf{r})\text{ in }\Omega
_{s}\\
0\text{ in }\overline{\Omega}_{sh}\\
\rho_{i}(\mathbf{r})=q_{i}\delta(\mathbf{r}-\mathbf{0})\text{ in }%
\overline{\Omega}_{i}%
\end{array}
\right.  ,\tag{4}\\
C_{k}(\mathbf{r}) &  =C_{k}^{\text{B}}\exp\left(  -\beta_{k}\phi
(\mathbf{r})+\frac{v_{k}}{v_{0}}S^{\text{trc}}(\mathbf{r})\right)  \text{ in
}\overline{\Omega}\text{,}\tag{5}\\
S^{\text{trc}}(\mathbf{r}) &  =\ln\left(  \frac{\Gamma(\mathbf{r)}}%
{\Gamma^{\text{B}}}\right)  \text{ in }\overline{\Omega}\text{,}\tag{6}%
\end{align}
where $\epsilon_{0}$ is the vacuum permittivity, $\epsilon_{\text{w}}$ is the
dielectric constant of bulk water, $\epsilon_{\text{ion}}$ is a dielectric
constant in $\overline{\Omega}_{i}$, $a_{j}$ is the radius of a counterion of
the ion $i$, and $\delta(\mathbf{r}-\mathbf{0})$ is the delta function at the origin.

The concentration function $C_{k}(\mathbf{r})$ is described by a Fermi
distribution (5), where $C_{k}^{\text{B}}$ is a constant bulk concentration
for all $k=1,\cdots,K+1$, $q_{K+1}=0$, $\beta_{k}=q_{k}/k_{B}T$, $v_{k}=4\pi
a_{k}^{3}/3$, $v_{0}=\left(  \sum_{k=1}^{K+1}v_{k}\right)  /(K+1)$ an average
volume of all kinds of hard spheres, $S^{\text{trc}}(\mathbf{r})$ is called
the steric potential, $\Gamma^{\text{B}}=1-\sum_{k=1}^{K+1}v_{k}%
C_{k}^{\text{B}}$ is a constant void fraction, $\Gamma(\mathbf{r)}%
=1-\sum_{k=1}^{K+1}v_{k}C_{k}(\mathbf{r})$ is a void fraction function, and
$K+1$ denotes water. The radii of $\Omega_{i}$ and the outer boundary of
$\Omega_{sh}$ are denoted by $R_{i}^{Born}$ and $R_{i}^{sh}$, respectively,
whose values will be determined by experimental data. It is natural to choose
the Born radius $R_{i}^{Born}$ (not the ionic radius $a_{i} $) as the radius
of $\Omega_{i}$ \cite{BC00}. We consider both first and second shells of the
ion \cite{RI13,MP11}.

The potential $\phi^{0}(\mathbf{r})$ (in Eq. (1)) of the ideal system is
obtained by setting $\rho_{s}(\mathbf{r})=0$ in (4), i.e., all particles in
$\Omega_{s}$ do not electrostatically interact with each other since $q_{k}=0$
for all $k$. The domain $\Omega$ is chosen to be sufficiently large so that
$\phi(\mathbf{r})=0$ on the boundary of the domain $\partial\Omega$. The ideal
potential $\phi^{0}(\mathbf{r})$ is then a constant, i.e., $\Delta G_{i}^{0}$
is a constant reference chemical potential independent of $C_{k}^{\text{B}}$.

The distribution (5) is of Fermi type since all concentration functions have
an upper bound, i.e., $C_{k}(\mathbf{r})<1/v_{k}$ for all particle species
with any arbitrary (or even infinite) potential $\phi(\mathbf{r})$ at any
location $\mathbf{r}$ in the domain $\Omega$ \cite{LE14}. The Poisson-Fermi
equation (2) and the Fermi distribution (5) reduce to the Poisson-Boltzmann
equation and the Boltzmann distribution when $l_{c}=S^{\text{trc}}=0$, i.e.,
when the correlation and steric effects are not considered. The Boltzmann
distribution $C_{k}(\mathbf{r})=C_{k}^{\text{B}}\exp\left(  -\beta_{k}%
\phi(\mathbf{r})\right)  $ would however diverge if $\phi(\mathbf{r})$ tends
to infinity. This is a major deficiency of PB theory for modeling a system
with strong local electric fields or interactions \cite{E12a}. If the
correlation length $l_{c}\neq0$, the dielectric operator $\widehat{\epsilon
}=\epsilon_{s}(1-l_{c}^{2}\nabla^{2})$ in Eq. (2) approximates the
permittivity of the bulk solvent and the linear response of correlated ions
\cite{S06,BS11,L13,LE13}, and yields a dielectric function $\widetilde
{\epsilon}(\mathbf{r})$ as an output of solving Eq. (2) \cite{LE14}. The exact
value of $\widetilde{\epsilon}(\mathbf{r})$ at any $\mathbf{r}\in$
$\overline{\Omega}_{sh}\cup\Omega_{s}$ cannot be obtained from Eq. (2) but can
be approximated by the simple formula $\widetilde{\epsilon}(\mathbf{r}%
)\approx\epsilon_{i}+$ $C_{\text{H}_{2}\text{O}}(\mathbf{r}\mathbb{)(}%
\epsilon_{s}- $ $\epsilon_{i})/C_{\text{H}_{2}\text{O}}^{\text{B}}$ since the
water density function $C_{\text{H}_{2}\text{O}}(\mathbf{r}\mathbb{)}%
=C_{K+1}(\mathbf{r}\mathbb{)}$ is an output of Eq. (5). This formula is only
for visualizing (approximately) the profile of $\widehat{\epsilon}$ or
$\widetilde{\epsilon}$. It is not an input of calculation. The input is the
correlation length $l_{c}$ in Eq. (3) \cite{S06,BS11,L13,LE13}. The actual
outputs are the numerical solutions of the partial differential equations and
boundary conditions.

The factor $v_{k}/v_{0}$ multiplying the steric potential function
$S^{\text{trc}}(\mathbf{r})$ in Eq. (5) is a modification of the unity used in
our previous work \cite{LE14,LE15a}. The steric energy $-\frac{v_{k}}{v_{0}%
}S^{\text{trc}}(\mathbf{r})k_{B}T$ \cite{LE14,LH16} of a type $k$ particle
depends not only on the voidness ($\Gamma(\mathbf{r)}$) (or equivalently
crowding) at $\mathbf{r}$ but also on the volume $v_{k}$ of the particle
itself. If all $v_{k}$ are equal (and thus $v_{k}=v_{0}$), then all particle
species at any location $\mathbf{r}\in\overline{\Omega}_{sh}\cup\Omega_{s}$
have the same steric energy, i.e., uniform particles are indistinguishable in
steric energy. The steric potential is a mean-field approximation of
Lennard-Jones (L-J) potentials that describe local variations of L-J distances
(and thus empty voids) between any pair of particles. L-J potentials are
highly oscillatory and extremely expensive and unstable to compute numerically
\cite{LE14}. Calculations that involve L-J potentials, or even truncated
versions of L-J potentials must be extensively checked to be sure that results
do not depend on irrelevant parameters.

\section{Methods}

To avoid large errors in approximation caused by the delta function
$\delta(\mathbf{r}-\mathbf{0})$ in (4), the potential function can be
decomposed as \cite{CL03,GY07,L13}%
\begin{equation}
\phi(\mathbf{r})=\left\{
\begin{array}
[c]{l}%
\widetilde{\phi}(\mathbf{r})+\phi^{\ast}(\mathbf{r})+\phi^{\text{L}%
}(\mathbf{r})\text{\ in }\Omega_{i}\\
\widetilde{\phi}(\mathbf{r})\text{ in }\overline{\Omega}_{sh}\cup\Omega_{s}%
\end{array}
\right.  \text{,}\tag{7}%
\end{equation}
where $\phi^{\ast}(\mathbf{r})=q_{i}/(4\pi\epsilon_{i}\left\vert
\mathbf{r-0}\right\vert )$ and $\widetilde{\phi}(\mathbf{r})$ is found by
solving%
\begin{align}
\left(  l_{c}^{2}\nabla^{2}-1\right)  \nabla\cdot\epsilon_{s}\nabla
\widetilde{\phi}(\mathbf{r})  & =\rho(\mathbf{r})\text{ in }\overline{\Omega
}_{sh}\cup\Omega_{s}\tag{8}\\
-\nabla\cdot\epsilon_{i}\nabla\widetilde{\phi}(\mathbf{r})  & =0\text{ \ in
}\Omega_{i}\tag{9}%
\end{align}
without the singular source term $\rho_{i}(\mathbf{r})=q_{i}\delta
(\mathbf{r}-\mathbf{0})$ and with the interface conditions%
\begin{equation}
\left\{
\begin{array}
[c]{l}%
\left[  \widetilde{\phi}(\mathbf{r})\right]  =0\\
\left[  \epsilon(\mathbf{r})\nabla\widetilde{\phi}(\mathbf{r})\cdot
\mathbf{n}\right]  =\epsilon_{i}\nabla\left(  \phi^{\ast}(\mathbf{r}%
)+\phi^{\text{L}}(\mathbf{r}))\right)  \cdot\mathbf{n}%
\end{array}
\right.  \text{ for all }\mathbf{r}\in\partial\Omega_{i}\text{,}\tag{10}%
\end{equation}
where $\mathbf{n}$ is an outward normal unit vector at $\mathbf{r}\in$
$\partial\Omega_{i}$ and the jump function $[u(\mathbf{r})]=\lim
_{\mathbf{r}_{sh}\rightarrow\mathbf{r}}u(\mathbf{r}_{sh})-\lim_{\mathbf{r}%
_{i}\rightarrow\mathbf{r}}u(\mathbf{r}_{i})$ with $\mathbf{r}_{sh}\in$
$\Omega_{sh}$ and $\mathbf{r}_{i}\in$ $\Omega_{i}$ \cite{L13}. The potential
function $\phi^{\text{L}}(\mathbf{r})$ is the solution of the Laplace
equation
\begin{equation}
\nabla^{2}\phi^{\text{L}}(\mathbf{r})=0\text{ in }\Omega_{i}\tag{11}%
\end{equation}
with the boundary condition%
\begin{equation}
\phi^{\text{L}}(\mathbf{r})=\phi^{\ast}(\mathbf{r})\text{ on }\partial
\Omega_{i}.\tag{12}%
\end{equation}
The evaluation of the Green's function $\phi^{\ast}(\mathbf{r})$ on
$\partial\Omega_{i}$ always yields finite numbers and thus avoids the
singularity in the solution process. The desired solvation energy $\Delta
G_{i}$ in Eq. (1) (and thus the individual ionic activity coefficient
$\gamma_{i}$) is then evaluated by \cite{GY07,L13}%
\begin{equation}
\Delta G_{i}=k_{B}T\ln\gamma_{i}=\frac{1}{2}q_{i}\left[  \widetilde{\phi
}(\mathbf{0})+\phi^{\text{L}}(\mathbf{0})\right]  \text{.}\tag{13}%
\end{equation}
Since the interface $\partial\Omega_{i}$ is a sphere centered at the origin,
the Laplace potential $\phi^{\text{L}}(\mathbf{r})=q_{i}/(4\pi\epsilon
_{i}R_{i}^{Born})$ is a constant in $\overline{\Omega}_{i}$, i.e., Eq. (11)
has been exactly solved.

The Poisson-Fermi equation (8) is a nonlinear fourth-order partial
differential equation (PDE) in $\Omega_{s}$. Newton's iterative method is
usually used for solving nonlinear problems. We seek a sequence of approximate
solutions $\left\{  \widetilde{\phi}_{m}(\mathbf{r})\right\}  _{m=1}^{M}$ by
iteratively solving the linearized PF equation%
\begin{equation}
\left(  l_{c}^{2}\nabla^{2}-1\right)  \nabla\cdot\epsilon\nabla\widetilde
{\phi}_{m}-\rho_{s}^{\prime}(\widetilde{\phi}_{m-1})\text{ }\widetilde{\phi
}_{m}=\rho_{s}(\widetilde{\phi}_{m-1})-\rho_{s}^{\prime}(\widetilde{\phi
}_{m-1})\text{ }\widetilde{\phi}_{m-1}\text{ in }\Omega_{s}\text{,}\tag{14}%
\end{equation}
until a tolerable potential function $\widetilde{\phi}_{M}$ is reached, where
$\widetilde{\phi}_{0}(\mathbf{r})$ is a given initial guess potential
function, $\rho_{s}(\widetilde{\phi}_{m-1})=\sum_{k=1}^{K}q_{k}C_{k}%
^{m-1}(\mathbf{r})$, $C_{k}^{m-1}(\mathbf{r})=C_{k}^{\text{B}}\exp\left(
-\beta_{k}\widetilde{\phi}_{m-1}(\mathbf{r})+\frac{v_{k}}{v_{0}}%
S_{m-1}^{\text{trc}}(\mathbf{r})\right)  $, $S_{m-1}^{\text{trc}}%
(\mathbf{r})=\ln\left(  \frac{\Gamma_{0}(\mathbf{r)}}{\Gamma^{\text{B}}%
}\right)  $, $\Gamma_{m-1}(\mathbf{r)}=1-\sum_{k=1}^{K+1}v_{k}C_{k}%
^{m-1}(\mathbf{r})$, $\rho_{s}^{\prime}(\widetilde{\phi}_{m-1})=\sum_{k=1}%
^{K}\left(  -\beta_{k}q_{k}\right)  C_{k}^{m-1}(\mathbf{r})$, and $\rho
_{s}^{\prime}(\widetilde{\phi})=\frac{d}{d\widetilde{\phi}}\rho_{s}%
(\widetilde{\phi})$. Note that the differentiation in $\rho_{s}^{\prime
}(\widetilde{\phi})$ is performed only with respect to $\widetilde{\phi}$
whereas $S^{\text{trc}} $ is treated as another independent variable although
$S^{\text{trc}}$ depends on $\widetilde{\phi}$ as well. Therefore, $\rho
_{s}^{\prime}(\widetilde{\phi})$ is not exact implying that this is an inexact
Newton's method \cite{DE82}.

The fourth-order problem can be resolved by transforming Eq. (14) into two
second-order PDEs \cite{L13}%
\begin{align}
\epsilon_{s}\left(  l_{c}^{2}\nabla^{2}-1\right)  \Psi(\mathbf{r}) &
=\rho(\widetilde{\phi}_{m-1})\text{ in }\overline{\Omega}_{sh}\cup\Omega
_{s}\tag{15}\\
-\epsilon_{s}\nabla^{2}\widetilde{\phi}_{m}(\mathbf{r})-\rho^{\prime
}(\widetilde{\phi}_{m-1})\text{ }\widetilde{\phi}_{m}(\mathbf{r}) &
=-\epsilon_{s}\Psi(\mathbf{r})-\rho^{\prime}(\widetilde{\phi}_{m-1})\text{
}\widetilde{\phi}_{m-1}\text{ in }\overline{\Omega}_{sh}\cup\Omega_{s}\tag{16}%
\end{align}
by introducing a density like variable $\Psi=\nabla^{2}$ $\widetilde{\phi} $
for which the boundary condition is \cite{L13}
\begin{equation}
\Psi(\mathbf{r})=0\text{ on }\partial\Omega_{s}\text{.}\tag{17}%
\end{equation}
Eqs. (9), (15), and (16) are coupled together in the entire domain $\Omega$
with the jump conditions in (10). Note that linear PDEs (14), (15), and (16)
converge to the nonlinear PDE (8) if $\widetilde{\phi}_{M}$ converges to the
exact solution $\widetilde{\phi}$ of Eq. (8) as $M\rightarrow\infty$, i.e.,
the approximate potential $\widetilde{\phi}_{M}(\mathbf{r})$ is sufficiently
close to the exact potential $\widetilde{\phi}(\mathbf{r})$ for all
$\mathbf{r}\in$ $\overline{\Omega}_{sh}\cup\Omega_{s}$ if the iteration number
$M$ is sufficiently large ($M\approx5$ to $37$ for this work with error
tolerance 10$^{-3}$).

The standard 7-point finite difference (FD) method is used to discretize all
PDEs (9), (15), and (16), where the jump conditions in (10) are handled by the
simplified matched interface and boundary (SMIB) method proposed in
\cite{L13}. For simplicity, the SMIB method is illustrated by the following 1D
linear Poisson equation (in $x$-axis)
\begin{equation}
-\frac{d}{dx}\left[  \epsilon(x)\frac{d}{dx}\widetilde{\phi}(x)\right]
=f(x)\text{ in }\Omega\tag{18}%
\end{equation}
with the jump condition%
\begin{equation}
\left[  \epsilon\widetilde{\phi}^{\prime}\right]  =-\epsilon_{i}\frac{d}%
{dx}\phi^{\ast}(x)\text{ at }x=\xi=\text{ }\partial\Omega_{i}\cap
\partial\Omega_{s}\text{,}\tag{19}%
\end{equation}
where $\Omega=$ $\Omega_{i}\cup\Omega_{s}$, $\Omega_{i}=(0,$ $\xi)$,
$\Omega_{s}=(\xi,$ $L)$,\ $f(x)=0$ in $\Omega_{i}$, $f(x)\neq0$ in $\Omega
_{s}$, and $\widetilde{\phi}^{\prime}=\frac{d}{dx}\widetilde{\phi}(x)$. The
corresponding cases to Eqs. (9), (15), and (16) in $y$- and $z$-axis follow in
a similar way. Let two FD grids points $x_{l}$ and $x_{l+1}$ across the
interface point $\xi$ be such that $x_{l}<\xi<x_{l+1}$ and $\xi=(x_{l}%
+x_{l+1})/2$ with $\Delta x=x_{l+1}-x_{l}=1$ \AA , a uniform mesh, for
example, as used in this work. The FD equations of the SMIB method at $x_{l}$
and $x_{l+1}$ are%
\begin{align}
\epsilon_{i}\frac{-\widetilde{\phi}_{l-1}+(2-c_{1})\widetilde{\phi}_{l}%
-c_{2}\widetilde{\phi}_{l+1}}{\Delta x^{2}}  & =f_{l}+\frac{c_{0}}{\Delta
x^{2}}\tag{20}\\
\epsilon_{s}\frac{-d_{1}\widetilde{\phi}_{l}+(2-d_{2})\widetilde{\phi}%
_{l+1}-\widetilde{\phi}_{l+2}}{\Delta x^{2}}  & =f_{l+1}+\frac{d_{0}}{\Delta
x^{2}},\tag{21}%
\end{align}
where%
\begin{align*}
c_{1}  & =\frac{\epsilon_{i}-\epsilon_{s}}{\epsilon_{i}+\epsilon_{s}}\text{,
}c_{2}=\frac{2\epsilon_{s}}{\epsilon_{i}+\epsilon_{s}}\text{, }c_{0}%
=\frac{-\epsilon_{i}\Delta x\left[  \epsilon\widetilde{\phi}^{\prime}\right]
}{\epsilon_{i}+\epsilon_{s}}\text{, }\\
d_{1}  & =\frac{2\epsilon_{i}}{\epsilon_{i}+\epsilon_{s}}\text{, }d_{2}%
=\frac{\epsilon_{s}-\epsilon_{i}}{\epsilon_{i}+\epsilon_{s}}\text{, }%
d_{0}=\frac{-\epsilon_{s}\Delta x\left[  \epsilon\widetilde{\phi}^{\prime
}\right]  }{\epsilon_{i}+\epsilon_{s}}\text{,}%
\end{align*}
$\widetilde{\phi}_{l}$ is an approximation of $\widetilde{\phi}(x_{l})$, and
$f_{l}=f(x_{l})$. Note that the jump value $\left[  \epsilon\widetilde{\phi
}^{\prime}\right]  $ at $\xi$ is calculated exactly since the derivative of
$\phi^{\ast}$ is given analytically.

Since the steric potential takes particle volumes and voids into account, the
shell volume $V_{sh}$ of the shell domain $\Omega_{sh}$ can be determined by
Eqs. (5) and (6) as%
\begin{equation}
S_{sh}^{\text{trc}}=\frac{v_{0}}{v_{\text{w}}}\ln\left(  \frac{O_{i}%
^{\text{w}}}{V_{sh}C_{K+1}^{\text{B}}}\right)  =\ln\left(  \frac
{V_{sh}-v_{\text{w}}O_{i}^{\text{w}}}{V_{sh}\Gamma^{\text{B}}}\right)
\text{,}\tag{22}%
\end{equation}
where the occupancy (coordination) number $O_{i}^{\text{w}}$ is given by
experimental data \cite{RI13,MP11}. The shell radius $R_{i}^{sh}$ of
$\Omega_{sh}$ is thus determined. Note that the shell volume depends not only
on $O_{i}^{\text{w}}$ but also on the bulk void fraction $\Gamma^{\text{B}}$,
namely, \textbf{on all salt and water concentrations} ($C_{k}^{\text{B}}$).

As discussed in \cite{VB15}, the solvation free energy of an ion $i$ should
vary with salt concentrations and can be expressed by a dielectric constant
$\epsilon(C_{i}^{\text{B}})$ that depends on the bulk concentration
$C_{i}^{\text{B}}$ of the ion. Therefore, the Born energy
\begin{equation}
\Delta G_{i}^{\text{Born}}=\left(  \frac{1}{\epsilon_{\text{w}}}-1\right)
\frac{q_{i}^{2}}{8\pi\epsilon_{0}R_{i}^{0}}\tag{23}%
\end{equation}
with the Born radius $R_{i}^{0}$ in pure water should be modified with the
concentration-dependent dielectric constant $\epsilon(C_{i}^{\text{B}})$.
Equivalently, the Born radius in electrolyte solutions can be modified from
$R_{i}^{0}$ by a simple formula
\begin{equation}
R_{i}^{Born}(C_{i}^{\text{B}})=\theta(C_{i}^{\text{B}})R_{i}^{0}\text{,
\ \ }\theta(C_{i}^{\text{B}})=\alpha_{1}^{i}+\alpha_{2}^{i}\left(
\overline{C}_{i}^{\text{B}}\right)  ^{1/2}+\alpha_{3}^{i}\left(  \overline
{C}_{i}^{\text{B}}\right)  ^{3/2}\text{,}\tag{24}%
\end{equation}
where $\overline{C}_{i}^{\text{B}}=$ $C_{i}^{\text{B}}$/M is a dimensionless
bulk concentration of type $i$ ions, M is the molar concentration unit, and
$\alpha_{1}^{i}$, $\alpha_{2}^{i}$, and $\alpha_{3}^{i}$ are adjustable
parameters for modifying the experimental Born radius $R_{i}^{0}$ to fit
experimental activity coefficients $\gamma_{i}$ that change with the bulk
concentration conditions $C_{i}^{\text{B}}$ of the ion. The Born radii
$R_{i}^{0}$ in Table 1 are cited from \cite{VB15}, which are computed from the
experimental hydration Helmholtz free energies of these ions given in
\cite{F04}. Numerical values in Tables 1 and 2 are all experimental data for
which their values are kept fixed throughout calculations once chosen.

The three parameters in Eq. (24) have physical or mathematical meanings unlike
many parameters in the Pitzer model \cite{F10}. Any model or numerical method
incurs errors to approximate a real system, i.e., it is impossible to obtain
real Born radius $R_{i}^{Born}(C_{i}^{\text{B}})$ exactly. The first parameter
$\alpha_{1}^{i}$ is an adjustment of the experimental Born radius $R_{i}^{0}$
when $C_{i}^{\text{B}}=0$ for all $i$. The second parameter $\alpha_{2}^{i}$
is an adjustment of $R_{i}^{Born}(C_{i}^{\text{B}})$ that accounts for the
real thickness of the ionic atmosphere (Debye length), which is proportional
to the square root of the ionic strength $\sqrt{I}$ in the Debye-H\"{u}ckel
theory \cite{LM03}. The third parameter $\alpha_{3}^{i}$ is simply an
adjustment in the next order approximation beyond the DH treatment of ionic atmosphere.

We summarize the mathematical solution process for determining the activity of
ionic solutions in the following algorithm.

\bigskip$\left.
\begin{array}
[c]{l}%
\text{1. Solve Eqs. (9), (10), and (16) for }\widetilde{\phi}\text{ with }%
\rho^{\prime}=\Psi=0\text{ (in pure water), }R_{i}^{Born}=R_{i}^{0}\text{,}\\
\text{ }\ \ \ \text{and }\phi^{\text{L}}=q_{i}/(4\pi\epsilon_{i}R_{i}%
^{0})\text{ to obtain }\Delta G_{i}^{0}\text{ by Eq. (13) and then set
}\widetilde{\phi}_{0}=\widetilde{\phi}\text{.}\\
\text{2. Solve Eqs. (15) and (17) for }\Psi\text{ with }R_{i}^{Born}\text{ in
(24).}\\
\text{3. Solve Eqs. (9), (10), and (16) for }\widetilde{\phi}_{m}\text{ with
}\phi^{\text{L}}=q_{i}/(4\pi\epsilon_{i}R_{i}^{Born})\text{ and then set
}\widetilde{\phi}_{m-1}=\widetilde{\phi}_{m}\text{. }\\
\text{ \ \ \ Go to 2 until convergence. }\\
\text{4. Obtain the activity coefficient }\gamma_{i}\text{ by Eq. (13).}%
\end{array}
\text{ }\right.  $

\begin{center}
$%
\begin{tabular}
[c]{c|c|c|c}%
\multicolumn{4}{c}{Table 1. Values of Model Notations}\\\hline
Symbol & Meaning & \ Value & \ Unit\\\hline
\multicolumn{1}{l|}{$k_{B}$} & \multicolumn{1}{|l|}{Boltzmann constant} &
\multicolumn{1}{|l|}{$1.38\times10^{-23}$} & J/K\\
\multicolumn{1}{l|}{$T$} & \multicolumn{1}{|l|}{temperature} &
\multicolumn{1}{|l|}{Table 2} & K\\
\multicolumn{1}{l|}{$e$} & \multicolumn{1}{|l|}{proton charge} &
\multicolumn{1}{|l|}{$1.602\times10^{-19}$} & C\\
\multicolumn{1}{l|}{$\epsilon_{0}$} & \multicolumn{1}{|l|}{permittivity of
vacuum} & \multicolumn{1}{|l|}{$8.85\times10^{-14}$} & F/cm\\
\multicolumn{1}{l|}{$\epsilon_{\text{ion}}$, $\epsilon_{\text{w}}$} &
\multicolumn{1}{|l|}{dielectric constants} & \multicolumn{1}{|l|}{$1$, Table
2} & \\
\multicolumn{1}{l|}{$l_{c}=2a_{j}$} & \multicolumn{1}{|l|}{correlation length}
& \multicolumn{1}{|l|}{$j=\text{Cl}^{-}$ etc.} & \AA \\
\multicolumn{1}{l|}{$O_{i}^{\text{w}}$} & \multicolumn{1}{|l|}{in Eq. (22)} &
\multicolumn{1}{|l|}{18 \cite{RI13,MP11}} & \\
\multicolumn{1}{l|}{$a_{\text{Li}^{+}}$, $a_{\text{Na}^{+}}$, $a_{\text{K}%
^{+}}$} & \multicolumn{1}{|l|}{radii} & \multicolumn{1}{|l|}{$0.6$, $0.95$,
$1.33$} & \AA \\
\multicolumn{1}{l|}{$a_{\text{Mg}^{2+}}$, $a_{\text{Ca}^{2+}}$, $a_{\text{Ba}%
^{2+}}$} & \multicolumn{1}{|l|}{radii} & \multicolumn{1}{|l|}{$0.65$, $0.99$,
$1.35$} & \AA \\
\multicolumn{1}{l|}{$a_{\text{F}^{-}}$,$a_{\text{Cl}^{-}}$, $a_{\text{Br}^{-}%
}$, $a_{\text{H}_{2}\text{O}}$} & \multicolumn{1}{|l|}{radii} &
\multicolumn{1}{|l|}{$1.36$, $1.81$, $1.95$, $1.4$} & \AA \\
\multicolumn{1}{l|}{$R_{\text{Li}^{+}}^{0}$, $R_{\text{Na}^{+}}^{0}$,
$R_{\text{K}^{+}}^{0}$} & \multicolumn{1}{|l|}{Born radii in Eq. (24)} &
\multicolumn{1}{|l|}{$1.3$, $1.618$, $1.95$} & \AA \\
\multicolumn{1}{l|}{$R_{\text{Mg}^{2+}}^{0}$, $R_{\text{Ca}^{2+}}^{0}$,
$R_{\text{Ba}^{2+}}^{0}$} & \multicolumn{1}{|l|}{Born radii} &
\multicolumn{1}{|l|}{$1.424$, $1.708$, $2.03$} & \AA \\
\multicolumn{1}{l|}{$R_{\text{Cl}^{-}}^{0}$, $R_{\text{Cl}^{-}}^{0}$,
$R_{\text{Cl}^{-}}^{0}$,} & \multicolumn{1}{|l|}{Born radii} &
\multicolumn{1}{|l|}{$1.6$, $2.266$, $2.47$} & \AA \\\hline
\end{tabular}
$\\[0pt]

\bigskip

$%
\begin{tabular}
[c]{crrrrrr}%
\multicolumn{7}{c}{Table 2. Values of $\epsilon_{\text{w}}$ at various $T$
\cite{FG97}.}\\\hline
$T$/K & 298.15 & 373.15 & 423.15 & 473.15 & 523.15 & 573.15\\
$\epsilon_{\text{w}}$ & 78.41 & 55.51 & 44.04 & 38.23 & 32.23 & 25.07\\\hline
\end{tabular}
$
\end{center}

\section{Results}

\begin{figure}[t]
\centering\includegraphics[scale=0.7]{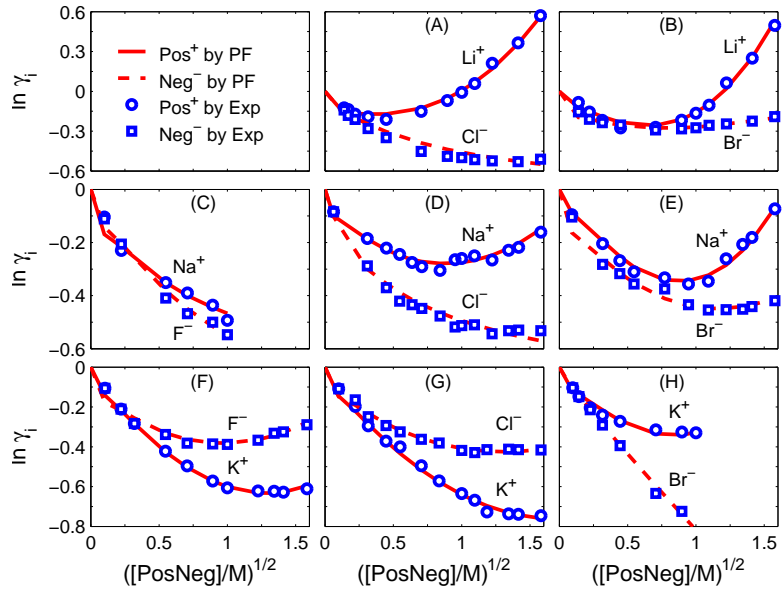}\caption{Indivivual activity
coefficients of 1:1 electrolytes. Comparison of PF results with experimental
data \cite{WR04} on $i=$ Pos$^{+}$ (cation) and Neg$^{-}$ (anion) activity
coefficients $\gamma_{i}$ in various [PosNeg] from 0 to 1.6 M.}%
\end{figure}

The PF results of ionic activity coefficients for eight 1:1 electrolytes, six
2:1 electrolytes, one mixed electrolyte, one 1:1 electrolyte at various
temperatures, and one 2:1 electrolyte at various temperatures agree with the
experimental data \cite{WR04,PP84,BH84,A92,GN78,WP98,C09,L65,KZ24,T21} as
shown in Figs. 2, 3, 4, 5, and 6, respectively. The empirical parameters used
to fit the experimental data are $\alpha_{1}^{i}$, $\alpha_{2}^{i}$, and
$\alpha_{3}^{i}$ in Eq. (24), whose values are given in Table 3 from which we
observe that the PF model requires only one to three parameters to fit those
data. \begin{figure}[t]
\centering\includegraphics[scale=0.7]{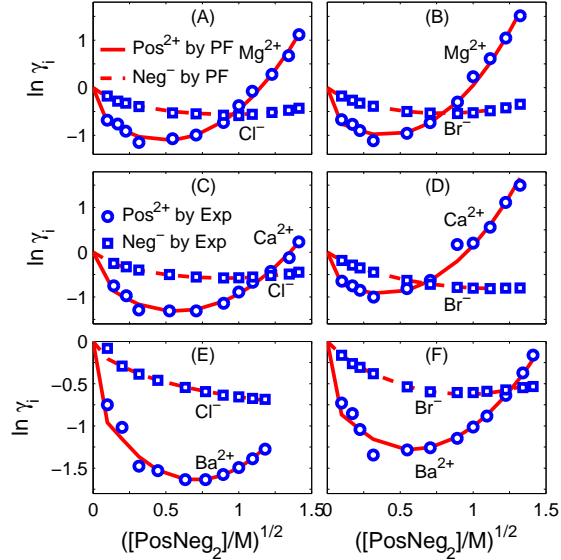}\caption{Indivivual activity
coefficients of 2:1 electrolytes. Comparison of PF results with experimental
data \cite{WR04} on $i=$ Pos$^{2+}$ (cation) and Neg$^{-}$ (anion) activity
coefficients $\gamma_{i}$ in various [PosNeg$_{2}$] from 0 to 1.5 M.}%
\end{figure}\begin{figure}[tt]
\centering\includegraphics[scale=0.7]{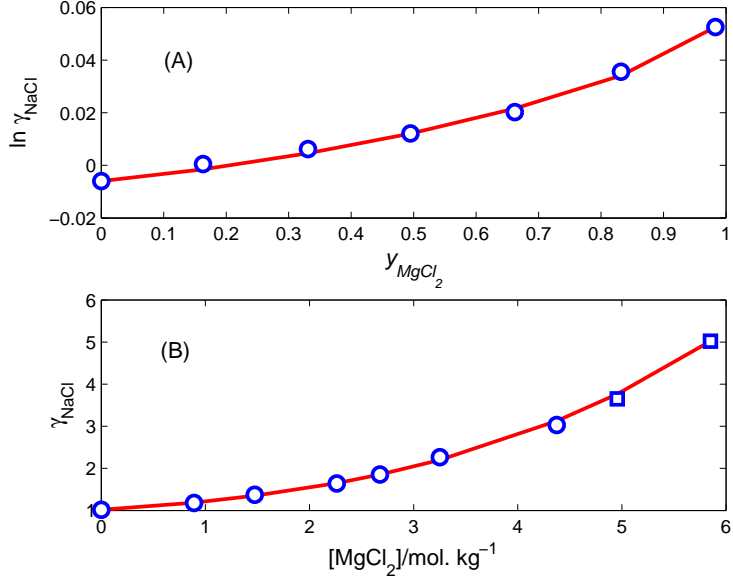}\caption{Mean activity
coefficients of mixed electrolytes. Comparison of PF results (curve) with
experimental data (symbols) compiled in \cite{RK15} (A) from \cite{L65} on
mean activity coefficients $\gamma$ of NaCl as a function of the ionic
strength ($I$) fraction $y_{\text{MgCl}_{2}}$of MgCl$_{2}$ in NaCl +
MgCl$_{2}$ mixtures at $I=$ 6 mol. kg$^{-1}$ and $T=$ 298.15 K; (B) from
\cite{KZ24} (circles)\ and \cite{T21} (squares) on $\gamma$ of NaCl as a
function of the MgCl$_{2}$ molality in NaCl + MgCl$_{2}$ mixtures at [NaCl]
$=$ 6 mol. kg$^{-1}$ and $T=$ 298.15 K.}%
\end{figure}\begin{figure}[ttt]
\centering\includegraphics[scale=0.7]{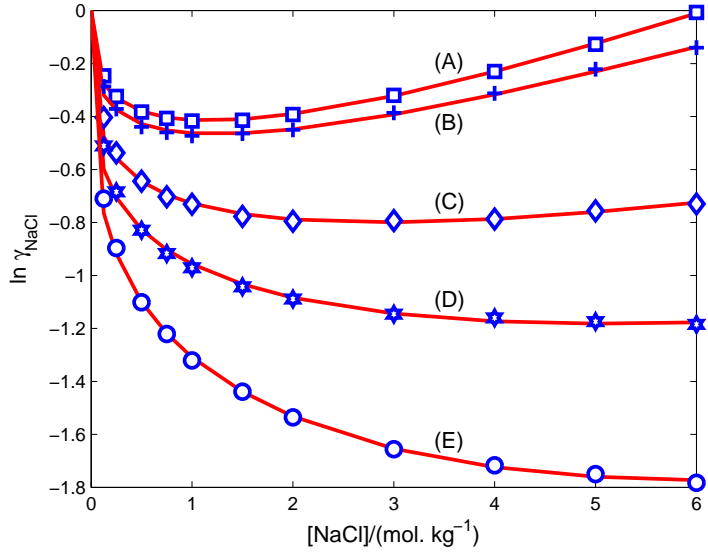}\caption{Mean activity
coefficients of 1:1 electrolyte at various temperatures. Comparison of PF
results (curves) with experimental data (symbols) compiled in \cite{RK15} from
\cite{PP84,BH84,A92} on mean activity coefficients $\gamma$ of NaCl in [NaCl]
from 0 to 6 mol. kg$^{-1}$ at $T=$ (A) 298.15 (B) 373.15 (C) 473.15 (D) 523.15
(E) 573.15 K.}%
\end{figure}\begin{figure}[tttt]
\centering\includegraphics[scale=0.7]{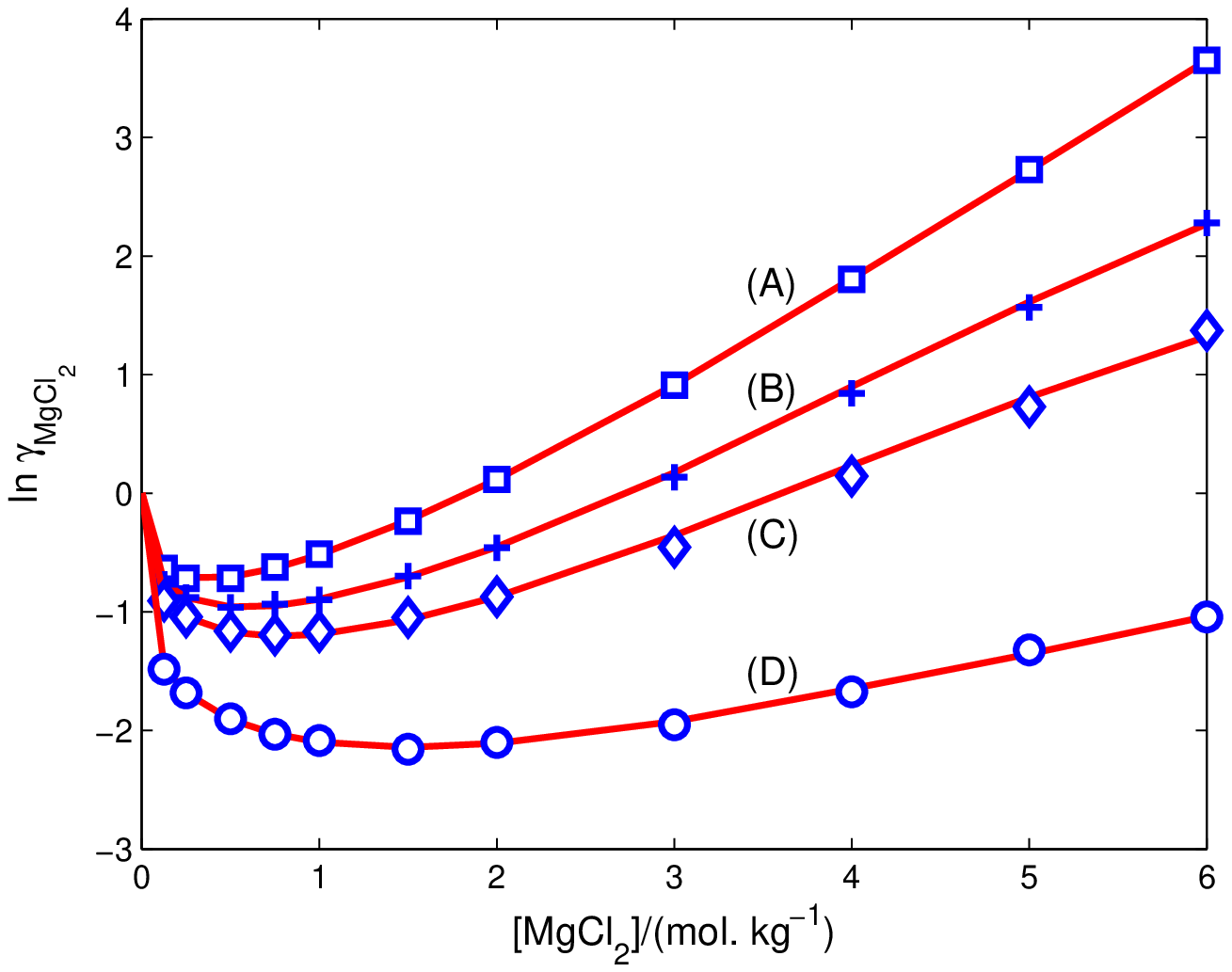}\caption{Mean activity
coefficients of 2:1 electrolyte at various temperatures. Comparison of PF
results (curves) with experimental data (symbols) compiled in \cite{RK15} from
\cite{GN78,WP98,C09} on mean activity coefficients $\gamma$ of MgCl$_{2}$ in
[MgCl$_{2}$] from 0 to 6 mol. kg$^{-1}$ at $T=$ (A) 298.15 (B) 373.15 (C)
423.15 (D) 523.15 K.}%
\end{figure}

The mean activity coefficient $\gamma_{PosNeg}$ of a salt Pos$_{p}$Neg$_{q}$
is calculated via the formula $\ln\gamma_{PosNeg}=\frac{p}{p+q}\ln\gamma
_{Pos}+\frac{q}{p+q}\ln\gamma_{Neg}$ \cite{LM03}, where $\gamma_{Pos}$ and
$\gamma_{Neg}$ are individual activity coefficients obtained by Eq. (13) for
each $i=Pos$ and $Neg$. For the mean activity coefficients of either ternary
(Fig. 4) or binary (Figs. 5 and 6) systems, we only need to adjust 3
parameters of one cation (not all ions) as shown in Table 3.

The activity coefficients by the PF model are quite successful over a large
range of temperatures and concentrations as shown in Figs. 4-6. We used the
code of the density model developed by Mao and Duan \cite{MD08} to convert the
concentration unit from molality (mol. kg$^{-1}$) to molarity (M = mol.
dm$^{-3}$) by the standard formula as given in \cite{MD08}, where the density
model has been compared with thousands of measurements at high accuracy. The
pressure values needed in the code at the corresponding temperatures were set
to $P=$ (A) 1.01 (B) 1.01 (C) 15.48 (D) 39.59 (E) 80.50 bar for Fig. 4 and (A)
1.01 (B) 1.01 (C) 4.73 (D) 39.50 bar for Fig. 5. In Fig. 6, the ionic strength
$I=\sum_{i}C_{i}^{\text{B}}z_{i}^{2}$ and the ionic strength fraction
$y_{\text{MgCl}_{2}}=$ $3m_{\text{MgCl}_{2}}/(3m_{\text{MgCl}_{2}%
}+m_{\text{NaCl}})$ with $m_{\text{MgCl}_{2}}$ and $m_{\text{NaCl}}$ being the
molalities of MgCl$_{2}$ and NaCl in the mixture, respectively, where $z_{i}$
is the valence of type $i$ ions.

We observe from Table 3 that the approximate $R_{i}^{Born}(C_{i}^{\text{B}})$
(with salts) deviates from $R_{i}^{0}$ (without salts) only in the second to
fourth decimal place, i.e., numerical values of $\gamma_{i}$ are very
sensitive to the decimal order of $\alpha_{1}^{i}$, $\alpha_{2}^{i}$, and
$\alpha_{3}^{i}$ because the Born radius $R_{i}^{Born}(C_{i}^{\text{B}})$ is
very close to the origin $\mathbf{0}$ at which the singular charge in
$\rho_{i}(\mathbf{r})=q_{i}\delta(\mathbf{r}-\mathbf{0})$ is infinite. The
approximation of the shell radius $R_{i}^{Sh}$ (or the coordination number
$O_{i}^{\text{w}}$ in Eq. (22)), on the other hand, is much less significant
than that of $R_{i}^{Born}$ because the electric potential $\phi^{\text{PF}%
}(\mathbf{r})$ diminishes exponentially in the hydration shell region
$\Omega_{sh}$ as shown by the profile of $\phi^{\text{PF}}(\mathbf{r})$ in
Fig. 7. The values of $\alpha_{1}^{i}$, $\alpha_{2}^{i}$, and $\alpha_{3}^{i}$
for each activity-concentration curve were obtained by first tuning three
values of $\theta(C_{i}^{\text{B}})$ in Eq. (24) to match three data points
($\sqrt{C_{ij}^{\text{B}}},$ $\ln\gamma_{ij}$) with three different
concentrations $C_{ij}^{\text{B}}$, $j=1,$ $2,$ $3,$ and then solving the
three unknowns $\alpha_{1}^{i}$, $\alpha_{2}^{i}$, and $\alpha_{3}^{i}$ using
three known $\theta(C_{ij}^{\text{B}})$ values. For example, for the $i=$
Li$^{+}$ curve in Fig. 2A, the selected experimental data points are
($\sqrt{C_{ij}^{\text{B}}},$ $\ln\gamma_{ij}$) = (0.315, -0.192), (1, -0.007),
(1.577, 0.57) and the corresponding tuned $\theta(C_{ij}^{\text{B}})$ are
0.9996, 1.0013, 1.0043.

The PF model can provide more physical details near the solvated ion
(Ca$^{2+}$, for example) in a strong electrolyte ([CaCl$_{2}$] = 2 M) such as
(1) the dielectric function $\widetilde{\epsilon}(\mathbf{r})$ with its
varying permittivity, (2) variable water density $C_{\text{H}_{\text{2}%
}\text{O}}(\mathbf{r})$, (3) concentration of counterion $C_{\text{Cl}^{-}%
}(\mathbf{r})$, (4) electric potential $\phi^{\text{PF}}(\mathbf{r})$, and (5)
the steric potential $S^{\text{trc}}(\mathbf{r})$ all shown in Fig. 7. The
steric potential is small because the configuration of particles (voids
between particles) does not vary too much from the solvated region to the bulk
region. Nevertheless, it has significant effect on the variation of mean-field
water densities $C_{\text{H}_{\text{2}}\text{O}}(\mathbf{r})$ and hence on the
dielectric function $\widetilde{\epsilon}(\mathbf{r})$ in the hydration
region. Note that $\widetilde{\epsilon}(\mathbf{r})$ is an output, not an
input of the model.

The strong electric potential $\phi^{\text{PF}}(\mathbf{r})$ in the Born
cavity $\overline{\Omega}_{i}$ (with $R_{i}^{Born}(C_{i}^{\text{B}})=1.7130$
\AA ) and the water density $C_{\text{H}_{\text{2}}\text{O}}(\mathbf{r})$ in
the hydration shell $\Omega_{sh}$ (with $R_{\text{Ca}^{2+}}^{sh}=5.0769$ \AA )
are the most important factors allowing the PF results to match the
experimental data. The ion and shell domains are the crucial region to study
ion activities. For example, Fraenkel's theory is entirely based on this
region --- the so-called smaller-ion shell region \cite{F10}. The steric
energy of water molecules modified by the factor $v_{K+1}/v_{0}$ in Eq. (5)
leads to significant changes of $C_{\text{H}_{\text{2}}\text{O}}(\mathbf{r})$
and $\widetilde{\epsilon}(\mathbf{r})$ profiles in Fig. 7 as compared with
those in Fig. 5 in our previous paper \cite{LE15a}.

\begin{center}
$%
\begin{tabular}
[c]{ccccclcccc}%
\multicolumn{10}{c}{Table 3. Values of $\alpha_{1}^{i}$, $\alpha_{2}^{i}$,
$\alpha_{3}^{i}$ in Eq. (24)}\\\hline
Fig.\# & $i$ & $\alpha_{1}^{i}$ & $\alpha_{2}^{i}$ & \ $\alpha_{3}^{i}$ &
\multicolumn{1}{|l}{Fig.\#} & $i$ & $\alpha_{1}^{i}$ & $\alpha_{2}^{i}$ &
\ $\alpha_{3}^{i}$\\\hline
\multicolumn{1}{l}{2A} & \multicolumn{1}{l}{Li$^{+}$} &
\multicolumn{1}{r}{$0.99913$} & \multicolumn{1}{r}{$0.00069$} &
\multicolumn{1}{r|}{$0.00009$} & 3C & \multicolumn{1}{l}{Ca$^{2+}$} &
\multicolumn{1}{r}{$0.99886$} & \multicolumn{1}{r}{$0.00046$} &
\multicolumn{1}{r}{$0.00011$}\\
\multicolumn{1}{l}{2A} & \multicolumn{1}{l}{Cl$^{-}$} &
\multicolumn{1}{r}{$0.99893$} & \multicolumn{1}{r}{$-0.00008$} &
\multicolumn{1}{r|}{} & 3C & \multicolumn{1}{l}{Cl$^{-}$} &
\multicolumn{1}{r}{$0.99877$} & \multicolumn{1}{r}{$-0.00060$} &
\multicolumn{1}{r}{$0.00012$}\\
\multicolumn{1}{l}{2B} & \multicolumn{1}{l}{Li$^{+}$} &
\multicolumn{1}{r}{$0.99958$} & \multicolumn{1}{r}{$-0.00019$} &
\multicolumn{1}{r|}{$0.00015$} & 3D & \multicolumn{1}{l}{Ca$^{2+}$} &
\multicolumn{1}{r}{$0.99886$} & \multicolumn{1}{r}{$0.00099$} &
\multicolumn{1}{r}{$0.00017$}\\
\multicolumn{1}{l}{2B} & \multicolumn{1}{l}{Br$^{-}$} &
\multicolumn{1}{r}{$0.99822$} & \multicolumn{1}{r}{$0.00107$} &
\multicolumn{1}{r}{} & \multicolumn{1}{|l}{3D} & \multicolumn{1}{l}{Br$^{-}$}
& \multicolumn{1}{r}{$0.99920$} & \multicolumn{1}{r}{$-0.00198$} &
\multicolumn{1}{r}{$0.00016$}\\
\multicolumn{1}{l}{2C} & \multicolumn{1}{l}{Na$^{+}$} &
\multicolumn{1}{r}{$0.99910$} & \multicolumn{1}{r}{} & \multicolumn{1}{r|}{} &
3E & \multicolumn{1}{l}{Ba$^{2+}$} & \multicolumn{1}{r}{$0.99844$} &
\multicolumn{1}{r}{$0.00011$} & \multicolumn{1}{r}{$0.00010$}\\
\multicolumn{1}{l}{2C} & \multicolumn{1}{l}{F$^{-}$} &
\multicolumn{1}{r}{$0.99933$} & \multicolumn{1}{r}{$-0.00029$} &
\multicolumn{1}{r|}{} & 3E & \multicolumn{1}{l}{Cl$^{-}$} &
\multicolumn{1}{r}{$0.99887$} & \multicolumn{1}{r}{$-0.00058$} &
\multicolumn{1}{r}{$0.00001$}\\
\multicolumn{1}{l}{2D} & \multicolumn{1}{l}{Na$^{+}$} &
\multicolumn{1}{r}{$0.99927$} & \multicolumn{1}{r}{$0.00026$} &
\multicolumn{1}{r|}{$0.00004$} & 3F & \multicolumn{1}{l}{Ba$^{2+}$} &
\multicolumn{1}{r}{$0.99851$} & \multicolumn{1}{r}{$0.00054$} &
\multicolumn{1}{r}{$0.00008$}\\
\multicolumn{1}{l}{2D} & \multicolumn{1}{l}{Cl$^{-}$} &
\multicolumn{1}{r}{$0.99840$} & \multicolumn{1}{r}{} & \multicolumn{1}{r|}{} &
3F & \multicolumn{1}{l}{Br$^{-}$} & \multicolumn{1}{r}{$0.99926$} &
\multicolumn{1}{r}{$-0.00145$} & \multicolumn{1}{r}{$0.00018$}\\
\multicolumn{1}{l}{2E} & \multicolumn{1}{l}{Na$^{+}$} &
\multicolumn{1}{r}{$0.99962$} & \multicolumn{1}{r}{$-0.00038$} &
\multicolumn{1}{r|}{$0.00010$} & 4A & \multicolumn{1}{l}{Na$^{+}$} &
\multicolumn{1}{r}{$1.00581$} & \multicolumn{1}{r}{$-0.00013$} &
\multicolumn{1}{r}{}\\
\multicolumn{1}{l}{2E} & \multicolumn{1}{l}{Br$^{-}$} &
\multicolumn{1}{r}{$0.99870$} & \multicolumn{1}{r}{$-0.00017$} &
\multicolumn{1}{r|}{$0.00004$} & 4B & \multicolumn{1}{l}{Na$^{+}$} &
\multicolumn{1}{r}{$1.00527$} & \multicolumn{1}{r}{$0.00042$} &
\multicolumn{1}{r}{$0.00019$}\\
\multicolumn{1}{l}{2F} & \multicolumn{1}{l}{K$^{+}$} &
\multicolumn{1}{r}{$0.99934$} & \multicolumn{1}{r}{$-0.00120$} &
\multicolumn{1}{r|}{$0.00007$} & 5A & \multicolumn{1}{l}{Na$^{+}$} &
\multicolumn{1}{r}{$0.9981$} & \multicolumn{1}{r}{} &
\multicolumn{1}{r}{$0.0001$}\\
\multicolumn{1}{l}{2F} & \multicolumn{1}{l}{F$^{-}$} &
\multicolumn{1}{r}{$0.99904$} & \multicolumn{1}{r}{$0.00013$} &
\multicolumn{1}{r|}{$0.00004$} & 5B & \multicolumn{1}{l}{Na$^{+}$} &
\multicolumn{1}{r}{$0.9971$} & \multicolumn{1}{r}{$0.0003$} &
\multicolumn{1}{r}{$0.0001$}\\
\multicolumn{1}{l}{2G} & \multicolumn{1}{l}{K$^{+}$} &
\multicolumn{1}{r}{$0.99929$} & \multicolumn{1}{r}{$-0.00122$} &
\multicolumn{1}{r|}{$0.00004$} & 5C & \multicolumn{1}{l}{Na$^{+}$} &
\multicolumn{1}{r}{$0.9945$} & \multicolumn{1}{r}{$-0.0007$} &
\multicolumn{1}{r}{$0.0001$}\\
\multicolumn{1}{l}{2G} & \multicolumn{1}{l}{Cl$^{-}$} &
\multicolumn{1}{r}{$0.99897$} & \multicolumn{1}{r}{$-0.00012$} &
\multicolumn{1}{r|}{$0.00003$} & 5D & \multicolumn{1}{l}{Na$^{+}$} &
\multicolumn{1}{r}{$0.9925$} & \multicolumn{1}{r}{$-0.0028$} &
\multicolumn{1}{r}{$0.0001$}\\
\multicolumn{1}{l}{2H} & \multicolumn{1}{l}{K$^{+}$} &
\multicolumn{1}{r}{$0.99931$} & \multicolumn{1}{r}{$0.00013$} &
\multicolumn{1}{r|}{} & 5E & \multicolumn{1}{l}{Na$^{+}$} &
\multicolumn{1}{r}{$0.9870$} & \multicolumn{1}{r}{$-0.0042$} &
\multicolumn{1}{r}{$0.0010$}\\
\multicolumn{1}{l}{2H} & \multicolumn{1}{l}{Br$^{-}$} &
\multicolumn{1}{r}{$0.99945$} & \multicolumn{1}{r}{$-0.00175$} &
\multicolumn{1}{r|}{$-0.00006$} & 6A & \multicolumn{1}{l}{Mg$^{2+}$} &
\multicolumn{1}{r}{$0.9988$} & \multicolumn{1}{r}{$0.0002$} &
\multicolumn{1}{r}{$0.0002$}\\
\multicolumn{1}{l}{3A} & \multicolumn{1}{l}{Mg$^{2+}$} &
\multicolumn{1}{r}{$0.99918$} & \multicolumn{1}{r}{$0.00044$} &
\multicolumn{1}{r|}{$0.00011$} & 6B & \multicolumn{1}{l}{Mg$^{2+}$} &
\multicolumn{1}{r}{$0.9989$} & \multicolumn{1}{r}{$-0.0004$} &
\multicolumn{1}{r}{$0.0003$}\\
\multicolumn{1}{l}{3A} & \multicolumn{1}{l}{Cl$^{-}$} &
\multicolumn{1}{r}{$0.99893$} & \multicolumn{1}{r}{$-0.00051$} &
\multicolumn{1}{r|}{$0.00010$} & 6C & \multicolumn{1}{l}{Mg$^{2+}$} &
\multicolumn{1}{r}{$0.9983$} & \multicolumn{1}{r}{$-0.0014$} &
\multicolumn{1}{r}{$0.0005$}\\
\multicolumn{1}{l}{3B} & \multicolumn{1}{l}{Mg$^{2+}$} &
\multicolumn{1}{r}{$0.99910$} & \multicolumn{1}{r}{$0.00063$} &
\multicolumn{1}{r|}{$0.00015$} & 6D & \multicolumn{1}{l}{Mg$^{2+}$} &
\multicolumn{1}{r}{$0.9961$} & \multicolumn{1}{r}{$-0.0020$} &
\multicolumn{1}{r}{$0.0003$}\\
\multicolumn{1}{l}{3B} & \multicolumn{1}{l}{Br$^{-}$} &
\multicolumn{1}{r}{$0.99888$} & \multicolumn{1}{r}{$-0.00065$} &
\multicolumn{1}{r|}{$0.00018$} &  & \multicolumn{1}{l}{} &
\multicolumn{1}{r}{} & \multicolumn{1}{r}{} & \multicolumn{1}{r}{}\\\hline
\multicolumn{10}{l}{Default values: $\alpha_{1}^{i}=1$, $\alpha_{2}^{i}=0$,
$\alpha_{3}^{i}=0$.}\\\hline
\end{tabular}
$
\end{center}

\begin{figure}[t]
\centering\includegraphics[scale=0.7]{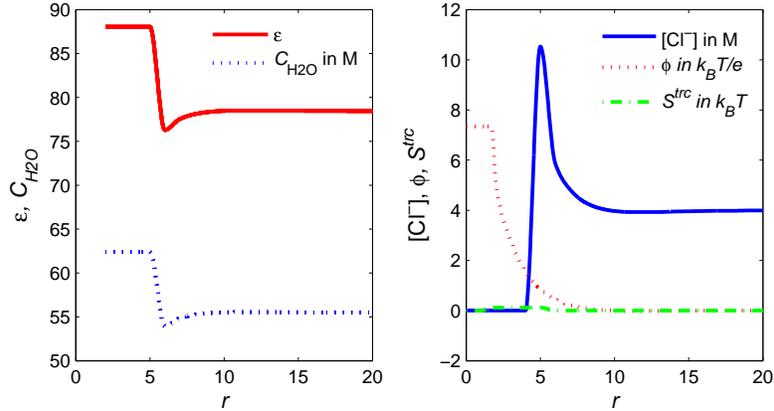}\caption{Dielectric function
$\widetilde{\epsilon}(\mathbf{r})$ (denoted by $\varepsilon$ in the figure),
water density $C_{\text{H}_{\text{2}}\text{O}}(\mathbf{r})$ ($C_{\text{H}%
_{\text{2}}\text{O}}$), Cl$^{-}$ concentration $C_{\text{Cl}^{-}}(\mathbf{r})$
([Cl$^{-}$]), electric potential $\phi^{\text{PF}}(\mathbf{r})$ ($\phi$), and
steric potential $S^{\text{trc}}(\mathbf{r})$ ($S^{\text{trc}}$) profiles near
the solvated ion Ca$^{2+}$ at [CaCl$_{2}$] $=2$ M, where $r$ is the distance
from the center of Ca$^{2+}$ in Angstrom.}%
\end{figure}

\section{Conclusion}

A Poisson-Fermi model for calculating activity coefficients of aqueous single
or mixed electrolyte solutions in a large range of concentrations and
temperatures has been presented and tested by a set of experimental data. The
model was shown to well fit experimental data with only three adjustable
parameters at most for each activity-concentration curve. The adjustable
parameters correspond to different orders of approximation of the unknown Born
radius of solvation energy that depends on salt concentrations in a highly
complex and nonlinear way. Nevertheless, the values of these parameters have
been shown to deviate slightly in decimal digits from that of the experimental
Born radius in pure water. These parameters are physically explained and can
be easily verified in future studies for the same or different solutions of
the present work. The model requires very few parameters because it is based
on an advanced continuum theory that accounts for steric and correlation
effects of ions and water with interstitial voids between nonuniform hard
spheres. It also deals with short- and long-range interactions by partitioning
the model domain into the ion, hydration shell, and the remaining solvent
sub-domains. Numerical methods were also given to show how to solve different
equations on different sub-domains that describe different physical properties
of an ion in electrolyte solutions.

\begin{acknowledgments}
This work was supported by the Ministry of Science and Technology, Taiwan
(MOST 105-2115-M-007-016-MY2 to J.L.L.).
\end{acknowledgments}

\end{document}